 \documentclass[final,3p,times,twocolumn]{elsarticle2}
\usepackage{amssymb}
\usepackage{graphicx}
\usepackage{bm}
\begin{document}

\begin{frontmatter}



\title{Crosstalk between DGP branes}


\author{Rainer Dick}
\ead{rainer.dick@usask.ca}
%%\author{A.B. Other}

\address{Department of Physics and Engineering Physics, 
University of Saskatchewan, Saskatoon, Canada SK S7N 5E2\\
and\\
Perimeter Institute for Theoretical Physics, 31 Caroline Street North, 
Waterloo, Canada ON N2L 2Y5}

\begin{abstract}
If two DGP branes carry $U(1)$ gauge theories and overlap, particles of one brane
can interact with the photons from the other brane. This coupling modifies in 
particular the Coulomb potentials between charges from the same brane
in the overlapping regions. The coupling also introduces Coulomb interactions
between charges from the different branes which can generate exotic bound states.

The effective modification of the fine structure constant in the overlap region
generates a trough in signals at the redshift of the overlap region and an
increase at smaller or larger redshift, depending on the value of the
crosstalk parameter $g_e g_p$. This implies potentially observable perturbations 
in the Lyman $\alpha$ forest if our 3-brane overlapped with another 3-brane in a 
region with redshift $z\lesssim 6$.  Crosstalk can also affect structure formation 
by enhancing or suppressing radiative cooling.
\end{abstract}

\begin{keyword}
Extensions of the Standard Model \sep Branes \sep Extra dimensions
 \sep Lyman $\alpha$ forest
\PACS 11.25.-w \sep 12.60.-i \sep 14.80.Rt 
\sep 98.58.Db \sep 98.62.Ra
%% MSC codes here, in the form: 
\MSC  83.E15 \sep 81.T30

\end{keyword}

\end{frontmatter}


\section{Introduction}
\label{sec:intro}
%%%%%%%%%%%%%%%%%%%%%%%%%%%%%%%%%%%%%%%%%%%%%%%%%%%%%%%%%%%%%%%%%%%%%%%%

The idea of extra dimensions has been around in theoretical physics for 
almost a century \cite{KK} and has been considerably expanded and 
reinvigorated in string theory. Furthermore, Dvali, Gabadadze and Porrati 
(DGP) pointed out in 2000 that we could live in a higher-dimensional world with
infinitely large extra dimensions hidden from plain sight because everything
except gravity can only propagate on a 3-brane in the higher-dimensional
world \cite{DGP,DG1}. The idea that observation of additional dimensions does 
not need to be suppressed by energy thresholds, but that instead there can 
be consistent restrictions of matter fields to submanifolds of a
higher-di\-men\-sio\-nal universe was a significant advancement 
of our understanding of higher dimensions. 
Therefore we denote a 3-brane carrying matter fields in an ambient spacetime
with gravitational degrees of freedom as a DGP brane, including also e.g.
3-branes in cascading gravity models \cite{cascade}. At this point we 
do not specify the background gravitational theory because we are interested
in electromagnetic effects on the branes.

Shortly after the inception of DGP branes, it was pointed out that at least
at the classical level they can support a modified Friedmann equation which 
may explain accelerated expansion without dark energy \cite{cedric,DDG}. 
Stability of the self-accelerated solution has meanwhile been called into 
question \cite{bend1}, but DGP branes can nevertheless support consistent 
modified cosmological evolution equations which comply with standard late 
time FLRW evolution \cite{cedric,DDG,rd,lue}. On the other hand, it was found 
in \cite{rd} and rediscovered in \cite{cordero} that DGP branes can 
even support the standard Friedmann equation and all the corresponding 
standard cosmological models on the brane, i.e. absence of cosmological 
signals from modified evolution equations does not rule out 
DGP branes. It is therefore important to also look
for other possible experimental signatures of DGP branes. 

In the present paper I would like to draw attention to the fact that
overlap of DGP branes at or after reionization can generate perturbations
in the Lyman $\alpha$ forest in the direction of the overlap region.
This is based on the observation that particles from our brane can
couple to photons from a $U(1)$ gauge theory on the second brane, thus 
impacting Coulomb interactions in the overlap region. This phenomenon of 
possible mixing of gauge interactions between two branes in an overlap 
region will be denoted as crosstalk. Indeed, it is possible and worthwhile 
to examine more general crosstalk models involving also Yukawa 
interactions between particles in overlapping brane volumes. We will focus 
on crosstalk interactions between charged particles and photons to study the 
impact of these interactions on electromagnetic potentials and the observed 
redshifts of spectral lines.
%%Baustelle!! description of sections

Crosstalk models, their impact on redshifts of spectral lines and 
consequences for the Lyman $\alpha$ forest are introduced in Sec. \ref{sec:z12}. 
Implications for structure formation and appearance of superlarge structures 
are oulined in Sec. \ref{sec:other}, and the conclusions are summarized 
in Sec. \ref{sec:conc}.

\section{Electromagnetic crosstalk between branes}
\label{sec:z12}

The action for fermions of masses $m_I$ and charges $q_I$ on our 3-brane is
\begin{eqnarray}\nonumber
S_1[\psi,A]\!\!\!\!&=&\!\!\!\!\int\!d^4x\,\mathcal{L}(\psi,A)
=\int\!d^4x\left[-\frac{1}{4}F^{\mu\nu}F_{\mu\nu}
\right.
\\ \label{eq:S1}
&&\!\!\!\!+\left.\sum_I\overline{\psi}_I\left(\mathrm{i}\gamma^\mu
\partial_\mu+q_I\gamma^\mu A_\mu-m_I\right)\psi_I
\right]
\end{eqnarray}
if we can neglect curvature effects. We make the same assumption of 
approximate flatness for the second brane.
At least weak curvature of at least one of the two branes will generically 
appear near the boundary of the overlap region, but we defer gravitational effects
of overlapping branes for later studies. Here we are primarily interested in the 
effects of electromagnetic crosstalk in approximately flat regions of overlap.

A simple example of smooth overlap of two flat 3-branes is e.g. provided by a flat
Minkowski 3-brane with inertial coordinates $x^\mu$ and perpendicular coordinate $\xi$
touched by a second brane with the same $t,y,z$ coordinates and the embedding
into flat ambient 5-dimensional spacetime given by
\begin{equation}\label{eq:touch1}
\xi=\frac{1}{3a^2}\Theta(x-a)(x-a)^3\pm\frac{1}{3a^2}\Theta(-x-a)(x+a)^3
\end{equation}
This 3-brane smoothly touches our Minkowski 3-brane for all values of $t,y,z$
and for $-a\le x\le a$. It actually smoothly penetrates through our 3-brane if 
we choose the plus sign in (\ref{eq:touch1}).

The induced metric on the second brane is
\begin{eqnarray*}
g_{\mu\nu}\!\!\!\!&=&\!\!\!\!\eta_{\mu\nu}+\frac{1}{a^4}\eta_\mu^1\eta_\nu^1\left[\Theta(x-a)(x-a)^4
\right.
\\
&&\!\!\!\!+\left.\Theta(-x-a)(x+a)^4\right],
\end{eqnarray*}
and vanishing of the Riemann tensor is easily verified. Of course, we can also
simply introduce inertial coordinates on the second brane by defining
$dX/dx=\sqrt{g_{xx}}$,
\begin{eqnarray*}
X\!\!\!\!&=&\!\!\!\!\frac{1}{a^2}\Theta(-x-a)\left[\int_0^{x+a}\!du\,\sqrt{a^4+u^4}-a^3\right]
\\
&&\!\!\!\!+\,\Theta(a^2-x^2)x
\\
&&\!\!\!\!+\,\frac{1}{a^2}\Theta(x-a)\left[\int_0^{x-a}\!du\,\sqrt{a^4+u^4}+a^3
\right].
\end{eqnarray*}

We are generically interested in finite volume overlaps, and then we do have 
to allow for at least weak curvature on the boundaries of the overlap region,
but this example and infinitely many similar examples demonstrate that 3-branes 
can smoothly overlap, share segments of their geodesics in the overlap region,
and yet be separately geodesically complete.  We will adapt the DGP framework
to the setting of smoothly overlapping 3-branes by postulating that each brane
carries its own field theory for the matter degrees of freedom, and that free
motion of those degrees of freedom corresponds to free fall along the geodesics 
in their own brane.
We do not allow for particle exchange between the branes, because in that case
we should rather consider a single brane having non-trivial topology and
carrying a single field theory for the matter degrees of freedom.

The second 3-brane will carry its own $U(1)$ gauge symmetry and charged particles 
with charges $\tilde{q}_J$ and masses $\tilde{m}_J$.  The corresponding fields 
are $\tilde{\psi}_J$ and $\tilde{A}_\mu$, and the corresponding action is 
\begin{equation}\label{eq:S2}
S_2[\tilde{\psi},\tilde{A}]=\int\!d^4\tilde{x}\,\mathcal{L}(\tilde{\psi},\tilde{A}).
\end{equation}

How could crosstalk work? The simplest (but still interesting) models would
assume Yukawa interactions involving scalar particles if we wish to stay within
the framework of renormalizable models. However, here we assume that our electrons 
and protons can see the photons from the second brane in those volumes where the 
branes overlap. Renormalizability implies that the coupling of the charged particles 
on our brane to the photons from the second brane in the overlap region $V_{12}$ is
\begin{equation}\label{eq:S12}
S_{12}[\psi,\tilde{A}]=\int\!dt\int_{V_{12}}\!d^3\bm{x}\sum_I g_I\overline{\psi}_I
\gamma^\mu\tilde{A}_\mu\psi_I,
\end{equation}
and a corresponding equation for the coupling $S_{21}[\tilde{\psi},A]$ 
of our photons to the fermions from the second brane. 
Note universality of the propagation speed of the
$U(1)$ gauge fields on both branes, because the free equations of motion for both
kinds of photons in the overlap region are
\[
\partial_\mu F^{\mu\nu}=0,\quad \partial_\mu\tilde{F}^{\mu\nu}=0.
\]

In the overlap region, the $U(1)$ from the second brane would enhance our
own $U(1)$ symmetry to $U(1)$ $\times$ $U(1)$,
\[
\psi'_I(x)=\exp\!\left(\mathrm{i}q_I f_1(x)
+\mathrm{i}g_I\tilde{f}(x)\right)\psi(x),
\]
\[
A'_\mu(x)=A_\mu(x)+\partial_\mu f(x),\quad
\tilde{A}'_\mu(x)=\tilde{A}_\mu(x)+\partial_\mu\tilde{f}(x).
\]
The onset of the additional $U(1)$ couplings at the boundary $\partial V_{12}$
of the overlap region generates steplike discontinuities in the equations
of motion but no $\delta$ function terms, since the discontinuties enter
only through the $\partial\mathcal{L}/\partial A_\mu$ 
and $\partial\mathcal{L}/\partial\overline{\psi}_I$ terms in
the Lagrange equations. 

Note that due to the lack of restrictions on $U(1)$ gauge couplings,
electromagnetic crosstalk between overlapping branes
appears like a natural and generic possibility if both branes carry
$U(1)$ gauge theories. The same cannot be said about non-abelian crosstalk,
since the gauge transformations for a non-abelian gauge field
\[
A'_\mu=U\cdot A_\mu\cdot U^{-1}+\frac{\mathrm{i}}{q}U\cdot\partial_\mu U^{-1}
\]
require universality of the gauge coupling $q$ for the non-abelian group.
Therefore, while the two branes would not need to carry the same sets of
representations of a non-abelian gauge group for corresponding crosstalk,
non-abelian crosstalk coupling constants would be restricted by the requirements
\begin{equation}\label{eq:nonabc1}
g=\tilde{q},\quad\tilde{g}=q.
\end{equation}

Calculating the energy-momentum tensor for $S=S_1+S_2+S_{12}+S_{21}$ 
in Coulomb gauge (see e.g. Sec. 21.4 in Ref. \cite{rdqm}) yields the Coulomb 
interaction terms in the Hamiltonian in the overlap region,
\begin{eqnarray}\nonumber
H_{11}\!\!\!\!&=&\!\!\!\!\sum_{II'}\left(q_I q_{I'}+g_I g_{I'}\right)
\int_{V_{12}}\!d^3\bm{x}\int\!d^3\bm{x}'
\\ \label{eq:H11}
&&\!\!\!\!\times\sum_{ss'}
\frac{\psi^+_{Is}(\bm{x})\psi^+_{I's'}(\bm{x}')\psi_{I's'}(\bm{x}')
\psi_{Is}(\bm{x})}{8\pi|\bm{x}-\bm{x}'|},
\end{eqnarray}
\begin{eqnarray}\nonumber
H_{12}\!\!\!\!&=&\!\!\!\!\sum_{IJ}\left(q_I\tilde{g}_{J}+g_I\tilde{q}_{J}\right)
\int_{V_{12}}\!d^3\bm{x}\int\!d^3\bm{x}'
\\ \label{eq:H12}
&&\!\!\!\!\times\sum_{ss'}
\frac{\psi^+_{Is}(\bm{x})\tilde{\psi}^+_{Js'}(\bm{x}')\tilde{\psi}_{Js'}(\bm{x}')
\psi_{Is}(\bm{x})}{4\pi|\bm{x}-\bm{x}'|},
\end{eqnarray}
and a corresponding term $H_{22}$ for the internal Coulomb interactions in the
second brane. Here we used the Schr\"o\-dinger picture field 
operators $\psi_{Is}(\bm{x})$ and $s,s'$ are Dirac labels.
Superficially, (\ref{eq:H11}) always looks repulsive between Dirac fields of
the same flavor, but recall that the actual particle and anti-particle creation
operators are $b^+_s(\bm{k})$ and $d^+_s(\bm{k})$, respectively. Substituting the 
mode expansions $\psi\sim b+d^+$ and normal ordering leads to the attractive 
Coulomb terms between particles and their anti-particles.

The effective modification of Coulomb interactions between charged particles
on our brane has all kinds of interesting possible consequences.
Everyday physics as we know it could be strongly modified in the overlap 
region. The electrostatic repulsion between electrons or protons would increase
according to $e^2\to e^2+g_e^2$ and $e^2\to e^2+g_p^2$, respectively.
The effective coupling constant between electrons
and protons would change from $-e^2$ to $-e^2+g_e g_p$. Hydrogen atoms could be 
weaker or more strongly bound, or not bound at all if
\begin{equation}\label{eq:ebarp}
-e^2+g_e g_p>0.
\end{equation}
In this case, positrons could bind with protons because charge conjugation 
still applies to the Dirac equations in the overlap region and 
therefore $g_{\overline{e}}=-g_e$.

The term (\ref{eq:H12}) would allow for the formation of exotic bound states 
of particles from the two branes. Furthermore, if we assume matter/anti-matter 
asymmetry also on the second brane, the Coulomb term (\ref{eq:H12}) seems 
to favor electromagnetic attraction between the branes if
\[
\sum_{IJ}\left(q_I\tilde{g}_{J}+g_I\tilde{q}_{J}\right)
=\sum_{I}g_I\sum_{J}\tilde{q}_{J}<0,
\]
and electromagnetic repulsion if $\sum_{I}g_I\sum_{J}\tilde{q}_{J}>0$. 
Here we used the fact that the sum over charges of non-confined low-energy 
particle states in our brane vanishes, $\sum_{I}q_I=q_e+q_p=0$. 

This leaves a lot of interesting possible implications of 3-brane overlap.
However, except for the particular case $e^2-g_e g_p=0$,
there will be hydrogen type bound states of particles with reduced 
mass $\mu=m_e m_p/(m_e+m_p)$ in the overlap region, albeit with a potentially 
very different effective fine structure constant
\[
\alpha_{12}=|e^2-g_e g_p|/4\pi.
\]
The energy levels of these hydrogen type atoms are therefore shifted
in leading order according to $E_{12,n}=(\alpha_{12}/\alpha)^2E_n$
which implies a corresponding shift in emitted or absorbed wavelengths
\begin{equation}\label{eq:lambda12}
\lambda_{12}=\frac{e^4}{(e^2-g_e g_p)^2}\lambda.
\end{equation}
The apparent redshift of the overlap region would therefore be
\begin{equation}\label{eq:z12}
z_{12}=(1+z)\frac{\lambda_{12}}{\lambda}-1
=\frac{z e^4+2e^2 g_e g_p-g_e^2 g_p^2}{(e^2-g_e g_p)^2},
\end{equation}
or in the case of very weak inter-brane gauge couplings, $|g_e g_p|\ll e^2$,
\[
z_{12}\simeq z+2(1+z)\frac{g_e g_p}{e^2}.
\]

If the second brane would carry a gauge group $U(1)^{\otimes n}$, the crosstalk 
parameter $g_e g_p$ would apparently have the form $\sum_{i=1}^n g^{(i)}_e g^{(i)}_p$.

We have
\begin{equation}\label{eq:ggcond}
z_{12}>z\,\,\Leftrightarrow\,\, 0<g_e g_p< 2e^2.
\end{equation}

The observational signature of a brane overlap region at a redshift $z\le 6$
would be a distortion of redshift binnings of hydrogen type clouds in the 
direction of the overlap region. For Lyman forests from quasars at $z\lesssim 6$ 
the signature would be a thinning of absorption lines in the range of the actual 
redshift parameter $z$ of the overlap region $V_{12}$, accompanied by higher
intensity of absorption lines at higher redshift or at lower redshift depending 
on whether the inequalities in (\ref{eq:ggcond}) hold or not. 
If the overlap region is near the onset 
of the Gunn-Peterson trough, it can delay or advance the apparent onset of the 
trough in the direction of $V_{12}$, i.e. reionization would appear to have occured 
earlier or later in the direction of $V_{12}$ than in other directions in our 3-brane.

\section{Other implications}
\label{sec:other}

Electromagnetic crosstalk increases repulsion between like particles and can weaken 
or strengthen electromagnetic coupling between electrons and protons depending 
on $g_e g_p$ (\ref{eq:ggcond}). This also implies that Bremsstrahlung 
emission into ordinary photons is weaker or stronger in $V_{12}$ since the emission 
probability will be proportional to $\alpha_{12}^2\alpha$. However, the total
Bremsstrahlung emission probability from electrons into both kinds of photons 
will be proportional to $\alpha_{12}^2[\alpha+(g_e^2/4\pi)]$, and the same proportionality 
also holds for dipole emission from atomic transitions.
This implies that we get weaker total electromagnetic emission in a smaller 
$g_e g_p$ range than the range in condition (\ref{eq:ggcond}),
\begin{eqnarray}\nonumber
&&\!\!\!\!\!\!\!\!\!\!\!\!P_{\gamma+\tilde{\gamma},12}<P_\gamma
\\ \label{eq:ggcond2}
&&\!\!\!\!\!\!\!\!\!\!\!\!\Leftrightarrow\,\,
e^2-\frac{e^3}{\sqrt{e^2+g_e^2}}<g_e g_p< e^2+\frac{e^3}{\sqrt{e^2+g_e^2}}.
\end{eqnarray}
We expect electromagnetic cooling of contracting gas clouds to be less 
efficient for $P_{\gamma+\tilde{\gamma},12}<P_\gamma$ and more efficient otherwise.
Increased mass density in an overlap region yields stronger curved geodesics,
but it does not help with the formation of stars and galaxies if cooling is 
suppressed. Therefore we would expect slower formation of stars and galaxies
in an overlap region where the inequalities in (\ref{eq:ggcond2}) hold, and
accelerated formation otherwise. The effect on structure formation 
should have the following consequences for the observed perturbation of 
absoption lines at the redshift $z$ of the region $V_{12}$:

If the inequalities in (\ref{eq:ggcond}) do not hold, the apparent 
redshift $z_{12}$ would satisfy $z_{12}<z$, and there could be more hydrogen 
clouds with higher column densities in the overlap region
due to $P_{\gamma+\tilde{\gamma},12}>P_\gamma$. The thinning out of 
absorption lines at $z$ should be there, but the increase at $z_{12}<z$
would be more pronounced than from the redshift effect (\ref{eq:z12}) alone.

On the other hand, if the inequalities in (\ref{eq:ggcond2}) hold, 
the apparent redshift $z_{12}$ would satisfy $z_{12}>z$, and there might 
also be fewer hydrogen clouds with smaller column densities in the overlap 
region. The thinning out of absorption lines at $z$ should be there, but the 
increase at $z_{12}>z$ would be less pronounced.
Please note that this scenario of reduced radiative cooling due to brane overlap 
could also help with the problem of overcooling in star formation histories, 
see e.g. \cite{mccarthy} and references there for a discussion of the overcooling 
problem.

As pointed out in Sec. \ref{sec:z12}, the primary observational effect of 
electromagnetic crosstalk between branes should be depletion of signals at 
the redshift $z$ of the overlap region $V_{12}$ and increase of signals at the
redshift $z_{12}$ (\ref{eq:z12}). Radiation sources in $V_{12}$ would then 
be assigned to higher or lower redshift values, depending on $g_e g_p$.
In terms of visible radiation signals, a large brane overlap region $V_{12}$ 
would then appear as a dark trough in front or behind an apparent wall,
or as a dark channel in front or behind an apparent filament.
Whether the observed superlarge structures at $z\sim 1.3$ \cite{Clowes} 
or $1.6<z<2.1$ \cite{Horvath} could be explained by brane crosstalk would 
then depend on successful correlation with corresponding perturbations
in the Lyman $\alpha$ forest. The discovery of these superlarge structures
could herald the dawn of brane astronomy.

\section{Conclusions}  
\label{sec:conc}

Equation (\ref{eq:H11}) shows that crosstalk between gauge theories in overlapping
branes affords local gauge couplings without promoting the couplings themselves to 
dynamical fields. This should impact the redshift distribution of Lyman $\alpha$ 
absorption lines by suppressing absorption lines at the redshift $z$ of the 
brane overlap region while increasing intensity of absorption lines at higher 
or lower redshift $z_{12}$ (\ref{eq:z12}), depending on the electromagnetic
crosstalk parameter $g_e g_p$.  The redshift distortion from overlapping
branes can also generate apparent large scale structure on scales that would 
violate size limits from structure formation in an isolated evolving 3-brane,
thus explaining the possible absence of an ``End of Greatness''.

It is known since 1971 that quasars shine light on the intergalactic medium.
Maybe quasars shine light on branes, too.

\section*{Acknowledgements}
This work was supported by NSERC Canada and by the Perimeter Institute for 
Theoretical Physics. Research at Perimeter Institute
is supported by the Government of Canada through Industry Canada and by the Province of
Ontario through the Ministry of Economic Development and Innovation.

\end{document}